# Asymmetric supercapacitors: optical and thermal effects when active carbon electrodes are embedded with nano-scale semiconductor dots

H. Grebel

The Center for Energy Efficiency, Resilience and Innovation (CEERI), The Electronic Imaging Center (EIC), The New Jersey Institute of Technology (NJIT), Newark, NJ 07102.
grebel@njit.edu

**Abstract:** Optical and thermal effects in asymmetric supercapacitors, whose active-carbon (AC) electrodes were embedded with nano-Si (n-Si) quantum dots (QD), are reported. We describe two structures: (1) p-n like, obtained by using a polyethylimine (PEI) binder for the 'n' electrode and a polyvinylpyrrolidone (PVP) binder for the 'p' electrode; (2) a single component binder – poly(methyl methacrylate) (PMMA). In general, AC appears black to the naked eye and one may assume that it acts as a black body absorber, namely, indiscriminately absorbing all light spectra. Yet, on top of a flat lossy spectra, AC (from two manufacturers) exhibited two distinct absorption bands: one in the blue (~400 nm) and the other one in the near IR (~840 nm). The n-Si material accentuated the absorption in the blue and bleached the IR absorption. Both bands contributed to capacitance increase: (a) when using aqueous solution and a PMMA binder, the optical related increased capacitance was 20% at low n-Si concentration and more than 100% for a high concentration dose; (b) when using IL electrolyte, the large, thermal capacitance increase (of ca 40%) was comparable to the optical effect (of ca 42%) and hence was assigned as an optically-induced thermal effect. The experimental data point to an optically induced capacitance increase even in the absence of the n-Si dots – this could be attributed to the absorption of AC in the blue.

## I. Introduction

Supercapacitors, S-C, [1-3] - capacitors that take advantage of the capacitance at the interface between an electrode and an electrolyte - have been found a large range of energy applications, least of all, optical modulators [4], and as buffering elements to subdue demand fluctuations in digital power networks [5-6]. A subclass of these are asymmetric S-C - capacitors made with two types of electrodes [7-9]. Asymmetric S-C are mostly fabricated as pseudo-supercapacitors - S-C whose capacitance is associated with a chemical reaction at the electrolyte/electrode interface. Pseudo-supercapacitors are aimed at increasing the operation potential range [10-14] via electrochemical means. To these, one ought to add dye sensitized solar cells [15-16], a special class of optically powered electrochemical energy sources with two distinct electrodes and an electrolyte that mediates the reacting ion species. Here, we describe carbon based, optically controlled asymmetric S-C that do not exhibit a chemical reaction at their electrolyte/electrode interfaces; namely, we focus on a basic, electrochemical double layer (EDL) supercapacitors. Our intent is to gain basic understanding of the optical and related thermal effects when incorporating n-Si dots in AC based electrodes. We describe two S-C types, which are both embedded with n-Si dots: polymeric doped AC electrodes via PVP and PEI binders, and non-doped AC electrodes with a PMMA binder.

## II. Methods and Experiments

The basic cell is composed of two transparent electrodes, either indium tin oxide (ITO, sheet resistance Rsqr=20 Ohms) or fluorinated doped tin oxide (FTO, sheet resistance Rsqr=30 Ohm) films on glass substrates. The electrodes were facing each other to make a parallel plate capacitor. The electrodes were coated with active carbon (AC) film (either produced by American Hardwood, AH, or by General Carbon Company, GCC). In the case of p-n cells, the AC was functionalized with binders using the same techniques used to functionalize carbon nanotubes [17-22]. While the AC is not a semiconductor material, nevertheless, it was hypothesized that the small AC domains would make it susceptible to polymeric doping.

**PVP and PEI binders, p-n like cells:** In the case of a p-n like cell, the AC (various concentrations in the range 100-200 mg/mL) in methanol was mixed using a sonicator with a horn antenna. The n-Si dots, of size less that 100 nanometers and at various concentrations in the range of 1-10 mg/mL were mixed in mostly the PVP 'p-type' material. The concentration range of the PVP was 20-40 mg/mL but larger than 10% by weight. Two molecular weights have been considered for the polymers: low m.w of 25 kD and large m.w of 630 kD. Two molecular weights were also considered for PEI: low m.w of 25 kD and 50% water diluted 600 kD. The low m.w polymers worked best. The irradiation on n-Si embedded PEI resulted in a reaction. The slurries were drop-casted on the transparent electrodes and were dried out at 90 °C for 30 min. The entire structure was held by a strong clip, leaving an exposed surface for light illumination. The illuminated (exposed) area was smaller than the entire area of the S-C. The films thickness varied but were of the order of a few hundreds of microns. As a result, capacitance of a given sample was assessed as a relative measurement: under light irradiation and without it.

**PMMA binder:** The adhesion of PEI and PVP to the conductive glass with IL was good but was sometimes compromised in aqueous cells. PMMM binder worked better under these circumstances with both IL and Na2SO4 electrolytes. The n-Si were typically incorporated within the electrode facing the light source. Typically, a 200 mg/mL AC produced by GCC with a 20 mg/mL PMMA binder on FTO was used.  The 2 mg/mL n-Si was mixed with the other components in Anisole and were dried out in an oven at 90 °C for 30 min.

**Electrolytes:** 98% Ionic liquid (IL, Alfa-Aesar) [23-25] and 1 M $Na_2SO_4$ were used as electrolytes.  For IL we used 1-n-Butyl-3-methyl-imidazolium hexafluro-phosphate.  The ionic liquid was soaking 0.1 mm thick lens tissues (Bausch & Lomb) used as separators.  Using hydrophilic nano-filtration polyamide filter (TS80, Sterlitech) proved to be too resistive for the as-is IL; both the tissue and the membrane worked well with the 1 M of $Na_2SO_4$ electrolyte.

**Electrochemical Measurements:** Measurements were carried out with a Potentiostat/Galvanostat (Metrohm).  The samples were irradiated with a 75 W incandescent light bulb situated 30 cm above the samples.  The light intensity of the entire radiation spectra (from the visible to the IR) was measured with a bolometer and was assessed as 30 $mW/cm^2$.  A calibrated homemade hot plate, which was interfaced with a thermocouple was used for the thermal experiments.  A second thermocouple assessed the temperature right at the sample surface.

**Electrical measurements:** Current-voltage plots (I-V curves) on dry films were obtained with a sensitive, 100 fA, computer controlled dedicated system (Keithley).

**Optical transmission measurements:** A computer controlled monochromator (SPEX), which was interfaced with a white light source, a chopper and a Si detector was used to assess the optical transmission of each film on a glass substrate.  The transmission value was assessed as the signal obtained with the film on the glass slide divided by the signal obtained with only the glass slide.

### III. Results and Discussion

*III.a. Film characterizations:*

**Asymmetric Cells:** The asymmetry of the capacitive element is demonstrated in Fig. 1.  Shown are cyclic voltammetry (CV) curves at a scan rate of 0.5 V/sec for a cell made with concentrations of AC, PVP, PEI, as follows: 100 mg/mL, 20 mg/mL and 20 mg/mL with lower m.w polymers.  The 'p'-like electrode was made of AC/PVP on Al and the 'n'-like electrode was made of AC/PEI also on Al.  The AC was made by AH and the electrolyte was IL soaking a tissue separator.  The curves are mirror image of each other when the positive and negative leads were switched, namely, the effect may not be attributed to Schottky barrier at the contacts.  Fig. 1a was obtained when the positive lead of the potentiostat was connect to AC/PVP electrode (denoted as $p^{(+)}$) and the negative lead of the potentiostat was connected to the AC/PEI side (denoted as $n^{(-)}$).  Fig. 1b was obtained when lead connections were switched - namely, $n^{(+)}$-$p^{(-)}$.

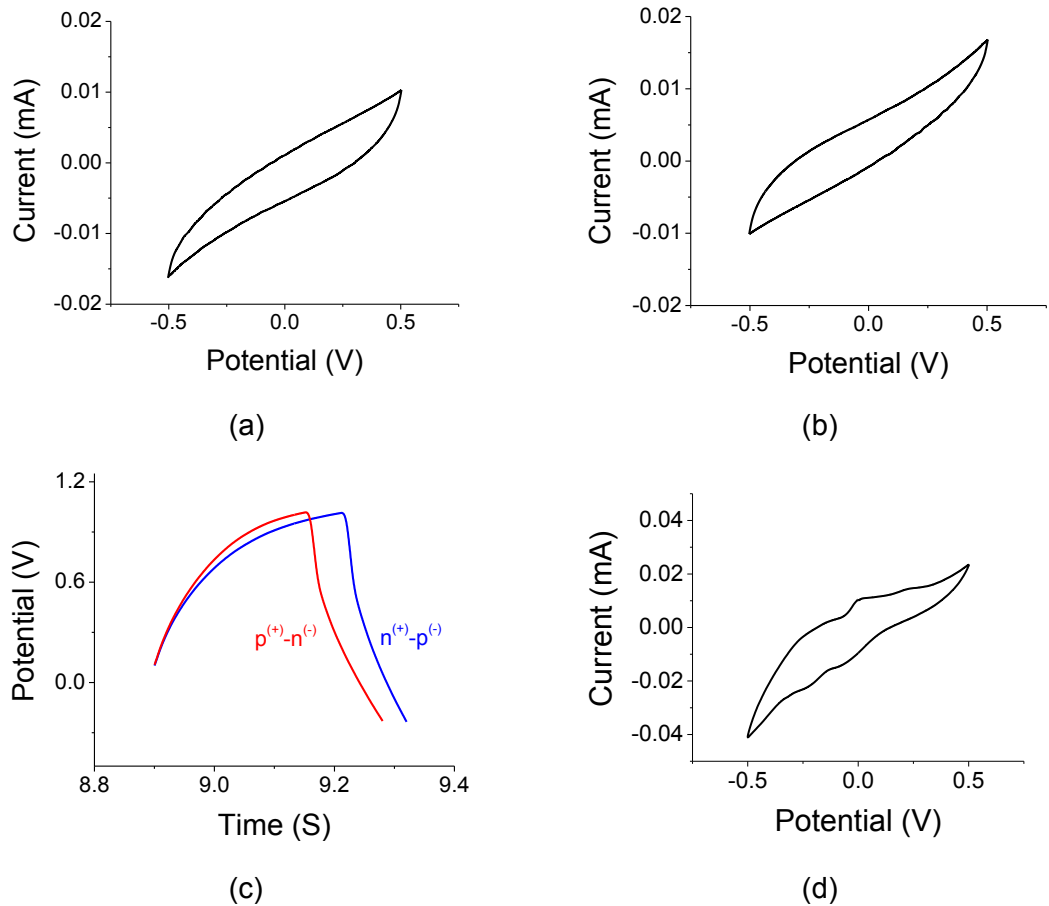

Fig. 1. (a,b) Cyclic voltammetry (CV) curves at scan rate of 0.1 V/s for AC/PVP-AC/PEI low m.w structure with IL electrolyte. (a) When the positive lead of the potentiostat was connected to the AC/PVP side and the negative lead was connected to the AC/PEI side. (b) When the leads have been switched: the positive lead of the potentiostat was connected to the AC/PEI side and the negative lead was connected to the AC/PVP side. (c) Charge-Discharge (CD) curves: red curve is when the positive lead is connect to AC/PVP; blue curve is when the positive lead is connected to AC/PEI. (d) Incorporating n-Si in both electrode material, and more specifically in AC/PEI resulted in a redox reaction while using an *aqueous* electrolyte.

Charge-discharge (CD) curves (Fig. 1c) conveyed the same message: the rise time is shorter and the discharge time is longer for $p^{(+)}$-$n^{(-)}$, whereas the reverse is true when the lead connections are switched. Granted that the S-C is less than perfect; it was of the order of tens of micro-F for this 0.5 cm$^2$ capacitor. Fig. 1d exhibits the redox reaction when the n-Si incorporated in the AC/PEI electrode.

**Optical Transmission and I-V curves:** The optical transmission through the various material components is shown in Fig. 2a. Since the films' thickness varied and we are only interested in the spectral shape of the curves, each transmission curve was normalized to its peak transmission. The glass slide signal was referenced to the transmission through air; all other curves were referenced to the signal of their substrate - a glass slide. The transmission of a glass slide is fairly constant throughout the spectral range between 400 to 900 nm. The transmission of the ITO and FTO on glass is flat throughout the visible with an overall transmission coefficient of 0.8; PMMA has also a flat transmission in the visible with a transmission coefficient of 0.9 (not shown).

The yellowish n-Si powder absorbs heavily in the blue green rang as typical for these nano-scale dots. The indirect bandgap of Si at 1.1 microns turns into a direct bandgap and its absorption is blue shifted when the dot size becomes smaller. The AC film exhibited two distinct absorption bands regardless of maker (the one depicted in the picture has been produced by American Hardwood (AH)). One absorption band is in the blue, centered at 460 nm and the other is in the near IR, centered at 840 nm. PVP absorbs in the deep blue and portrays flat absorption for wavelengths between 500 to 900 nm. The absorption of AC/PVP followed that of only AC. *Most notably is the transmission of PVP/n-Si (not shown) and AC/PVP/n-Si. The absorption peak at 840 nm disappeared, leaving only an absorption peak in the blue. Also noted is the blue shift for the 460 nm absorption line to below 400 nm with a combined effect that is larger than for either components.* The behavior at 840 nm may be explained if attribute the absorption line to impurity doping or surface states. Electrons were transferred from the AC (donors) to the n-Si (acceptors) and the transition was bleached. The behavior near 400 nm is more complex and could involve dipole coupling between the AC and n-Si species.

In Fig. 2b we show the I-V curve of illuminated and non-illuminated *dry* AC based films on ITO. A 1-mm scratch was made in the ITO layer, preventing electrical conduction across it (Fig. 2c). The film was bridging the gap and enabled conduction. There are two takeaways from the experiments: (1) The ITO/AC films exhibited Ohmic contacts; and (2) the film conductivity has increased under white light illumination. The films' thicknesses and widths were not the same, which explains the difference in the curves slopes. Nevertheless, we can assess the relative conductivity ($\sigma=I/V$) change under illumination: it was ($\sigma_{illuminated}-\sigma_{dark}$)/$\sigma_{dark}$=5.5x10$^{-2}$=5.5% for the AC/PVP/n-Si and 3x10$^{-2}$=3% for the AC/PVP film. As expected, the n-Si has improved the film conductivity under light even though its concentration was of the order of a few percent compared with a typical 20% conductive additives in commercial S-C material [25].

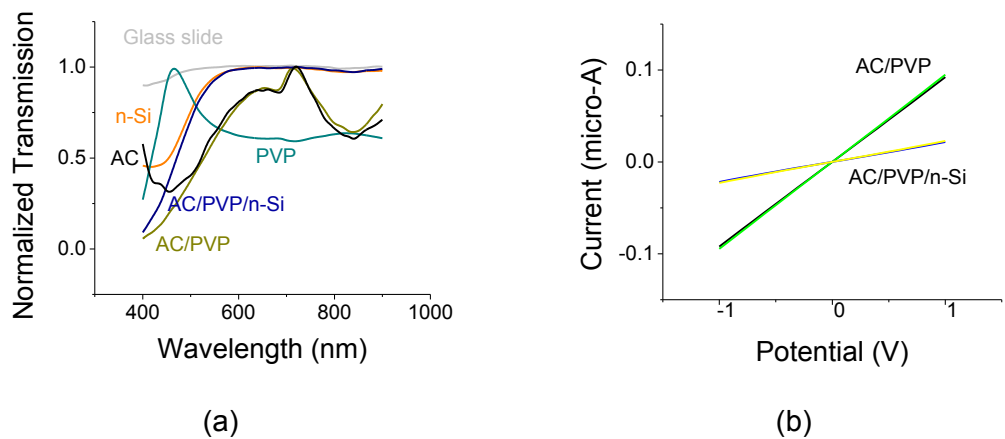

(a)  (b)

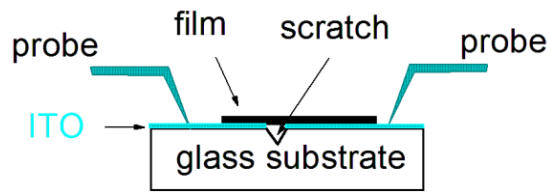

(c)

Fig. 2. (a) Normalized optical transmission through the various film components used in the experiments. The peak transmission was set to 1 for each curve in order to accentuate the spectral response. The transmission of the glass slide was referenced to air. All other transmission values were referenced to the transmitted signal through the glass slide. (b) I-V curves for AC/PVP and AC/PVP/n-Si under illumination of 30 mW/cm$^2$ light bulb and in the dark (room lighting). The inset shows the measurement arrangement. The effect is small but nonetheless measureable. (c) Measuring conductivity with and without illumination (impinging the film from above).

### *III.b. Supercapacitors under light ON and light OFF conditions*

In the remainder of this paper we will sort out the optical and related thermal effects in these films. In Fig. 3 we show CV and CD curves for a p-n cell - p(AC/PVP/n-Si)-n(AC/PEI) - under light ON and light OFF. The CV curves are a-symmetric when light is OFF (room light) and ON. The sample was deposited on ITO with lens tissue, serving as a separator and soaked with IL. Capacitance increase can be observed in Fig. 3a. The relative capacitance increase which should be normalized by the illuminated area $[(C/A)_{illum}-(C/A)_{dark}]/(C/A)_{dark}=0.42$ or a 42% increase under illumination. Upon illumination, one may observe a tilt of the plot axis towards larger current values. Such tilt may be attributed to a larger film conductivity and a lower ESR. The latter is corroborated by CD data (Fig. 3b). Capacitance here is calculated for the decay branch as $C_{eff} \sim (I_0/(V_0/\tau)$ which replaces the usual linear expression, $C=I_0/(dV/dt)$. Here $I_0$ is the (constant) discharge current, $1/\tau$ is a single decay rate that approximates the discharged branch, $V_0$ is the voltage difference ($V_0$=1 V in our case) and $C_{eff}$, the effective capacitance across the various film regions as it is gradually discharged. The relative capacitance change under illumination is ~37%. Overall, the ESR is quite large and may be attributed to the large impedance of the current collector (ITO; Rsqr=20 Ohms compared to 2.7 mOhms of Al), use of IL, and non-optimized binder to AC ratio.

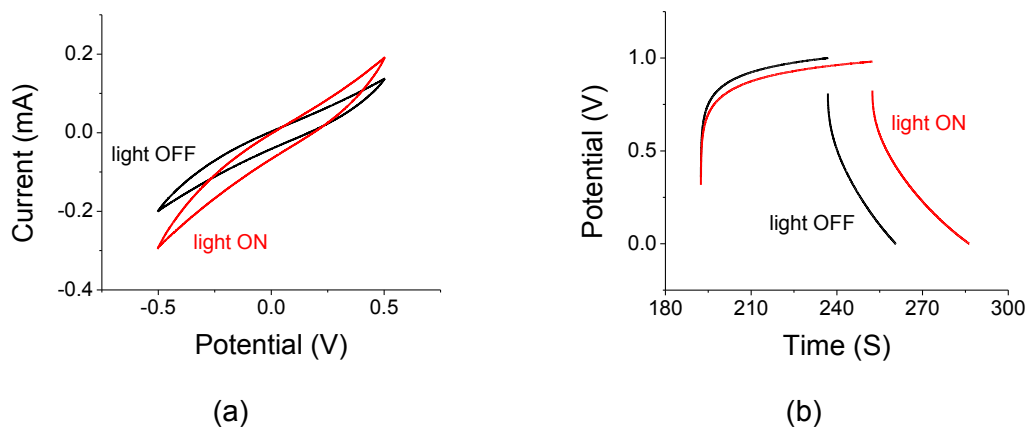

Fig. 3. (a) CV at a scan rate of 0.5 V/s under dark and illuminated conditions for a p(AC/PVP/n-Si)-n(AC/PEI) cell. (b) Corresponding CD curves. A small ESR decrease is noted under illumination. Note the rotated curve in (a) and the increase in both the charge and discharge time constants in (b) under the illumination with white light.

### *III.c. Thermal considerations during light ON and light OFF conditions*

### *III.c1: PMMA binder with aqueous solution*

Let us start with the simpler system using a single electrode binder (PMMA), where the electrode facing the light source is embedded with n-Si dots.   In Fig. 4a-b we present two CV plots: Fig. 4a has been obtained when heating the sample from 23 $^{o}$C to 35 $^{o}$C, at the rate of 0.1 $^{o}$C/sec, while collecting the CV data continuously.  The light was OFF.  The AC manufacture was GCC and the electrolyte was 1 M $Na_2SO_4$.  The loading of the n-Si was 2 mg/mL.  Note the rotation of the curve as the sample heats up; its waist at zero potential has changed only little.  This is our reference.  The relative change from the first scan to the last was <2%.  Fig. 4b exhibits two CV curves for light OFF and light ON.  Upon illumination, the temperature at the sample surface has elevated to 27 $^{o}$C.  The relative capacitance change (including the fact the exposed illuminated area is smaller than the area of the entire S-C) for the ON/OFF case was 20%, clearly, larger than the thermal reference.  While the curve has rotated, its waist has increased too.  Finally, CV curves were obtained with AC electrodes and only PMMA as a binder (namely, without n-Si).  As seen from Fig. 4c, there is a little optical effect when considering the smaller illuminated area of the S-C, while the temperature at the sample surface has reached 30 $^{o}$C under optical illumination.  Heating of the AC electrode even without n-Si is attributed to an overall AC scattering/absorption.  Thus, while the optical related heating is substantial, the increase in local polarization is small.

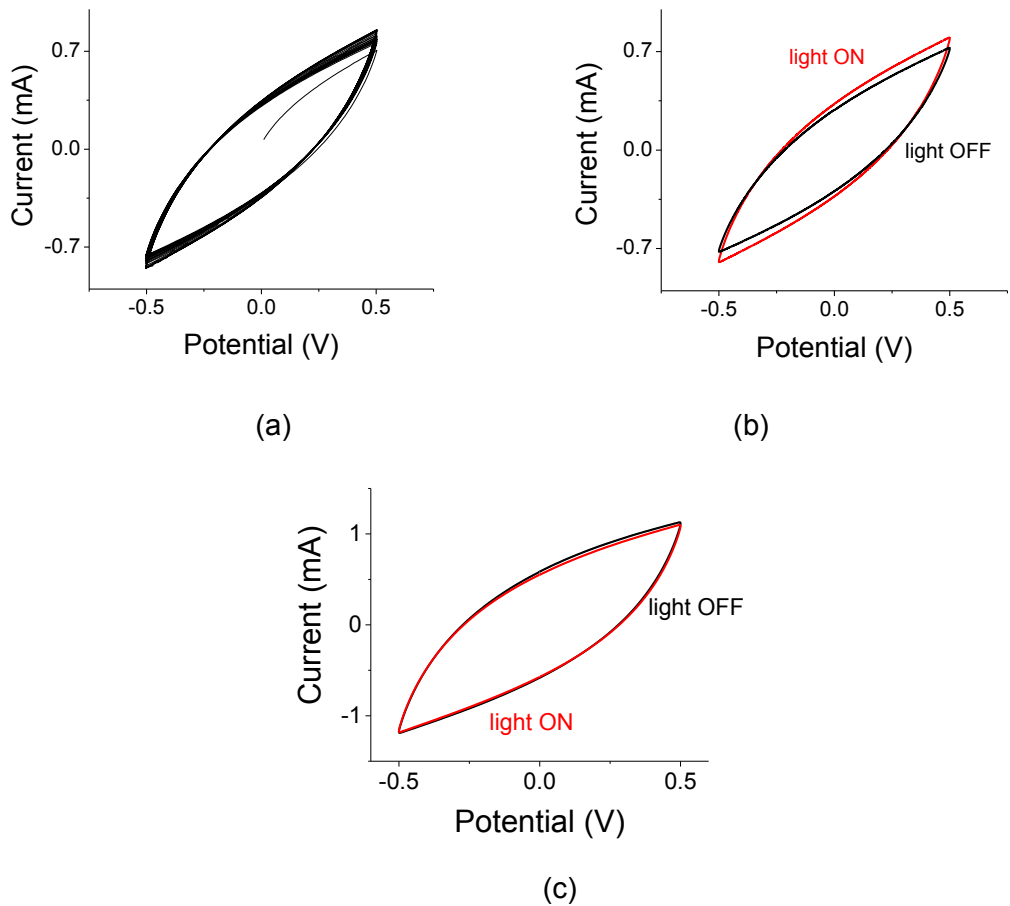

Fig. 4. CV curves for AC in PMMA binder with 1 M $Na_2SO_4$ electrolyte soaking a TS80 membrane and n-Si loading of 2mg/mL. Curves were obtained at a scan of 0.1 V/s. (a) Calibration curve when the sample was heated when light was OFF from 23 to 35 °C. Note the rotation of the curve as the sample heats up; its waist, however, at zero potential barely changed. (b) CV curves for light OFF (room lighting) and light ON. While the curve has rotated its waist has increased too. (c) AC with PMMA binder, yet without n-Si has exhibited a small optical effect if we take into account the smaller light illuminated exposed area.

An easy calibration method may be made with the one S-C cell. In that case one compares the effect of illumination for two cases: the case when the electrode containing n-Si dots is facing the light source and the case when the other electrode, the one without the dots is facing the light source. This experiment is presented in Fig. 5. Both electrodes were deposited on an ITO with Rsqr=5 Ohms (Huanyu). The electrolyte was $Na_2SO_4$ and the cell has reached 26°C in both cases during illumination. The concentration of the n-Si dots was 10 mg/mL, substantially larger than all other cases described before. Specifically, when electrode without n-Si was facing the light source, the relative OFF/ON change was ca 11% (22% when considering the smaller light exposed area); it was 56% (112% when considering the smaller light exposed area) for electrode containing the n-Si dot and facing the light source. Based on Fig. 4c and 5a, one may postulate that there could be an optical effect even in the absence of n-Si dots due to absorption of AC in the blue.

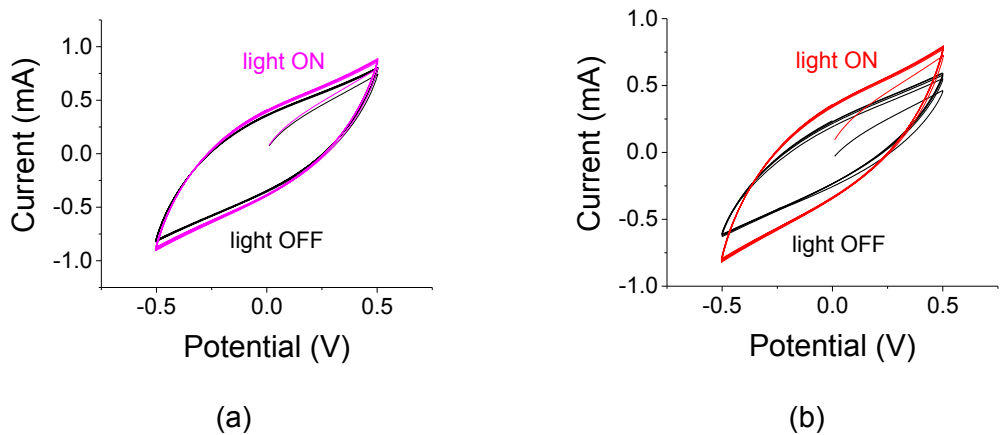

(a)                            (b)

Fig. 5. Measuring the cell capacitance when the electrode without the n-Si dot is facing the light source (a) and when the electrode containing the dots is facing the light source. In both cases, the surface of the sample reached 26°C. The n-Si loading was 10 mg/mL.

### *III.c2: PMMA binder with ion liquid electrolyte*

In order to sort out whether the capacitance increase is due to resulting heating of the electrolyte through light absorption, we repeated the previous measurements and replaced the aqueous solution with an IL electrolyte. The AC electrode here was embedded with a higher n-Si concentration - 10 mg/mL. Fig. 6a presents the capacitance increase upon heating from 23 to 45 °C at a rate of 0.1 °C/s. The CV data was continuously collected throughout the heating process. Fig. 6b are the related CV curves upon light ON and light OFF. A large capacitance increase is noted: the relative capacitance has increased by 50% when the sample was heating up from 23 °C to 37 °C (Fig. 6a). Similarly, a ca 50% capacitance increase and related temperature increase is noted based on Fig. 6b. Thus, unlike Fig. 4, here we cannot separate the optical effect from the related thermal effect. Similar conclusions may be drawn when a lower concentration of n-Si was used.

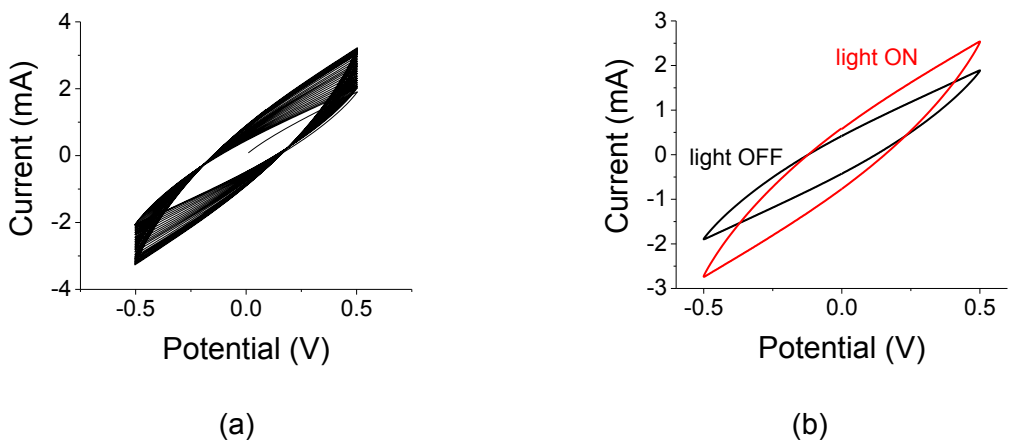

(a)                          (b)

Fig. 6. CV curves for AC embedded with high-concentration of n-Si (10 mg/mL) with a PMMA binder. The IL electrolyte was soaking a lens tissue. Curves were obtained at a scan of 0.1 V/s. (a) Calibration curve when the sample was heated when light was OFF from 23 to 45 °C. Note the rotation of the curve as the sample heats up. (b) CV curves with light OFF (room lighting) and light ON. The sample heated up from 23 °C to 37 °C upon illumination.

### *III.c3: p-n binders with ion liquid electrolyte*

CV were obtained for p-n like cell (n-Si in AC/PVP and AC/PEI, respectively. The electrolyte was IL. In Fig. 7a we present a CV curve when heating the sample from 23 °C to 35 °C while collecting the CV data continuously, all awhile the light source was OFF. The heating rate was 0.1 °C/sec. The relative change from the first scan to the last was ca 40%. Fig. 6b exhibits two CV curves for light OFF and light ON. Upon illumination, the temperature at the sample surface has elevated to 30 °C. The relative capacitance change for the ON/OFF case was 35%. Using the calibration curve of Fig. 7a, the thermal effect between 23-30 °C may account for only 23% of the capacitance increase while the remaining 13% could be attributed to the optical effect. This last value is similar to the value obtained for the aqueous system. Nonetheless, since the p-n is a more complex system than the one involving PMMA, and recognizing that the conductivity of PVP changes upon light illumination (Fig. 2b), more data are needed to affirmatively determine the portion of the optical effect with IL electrolytes. We point to recent studies that showed that the thermal sensitivity of IL could be substantially decrease by using a proper mixtures [24] and could be used here to accentuate the optical effect; such a study is beyond the scope of the current manuscript.

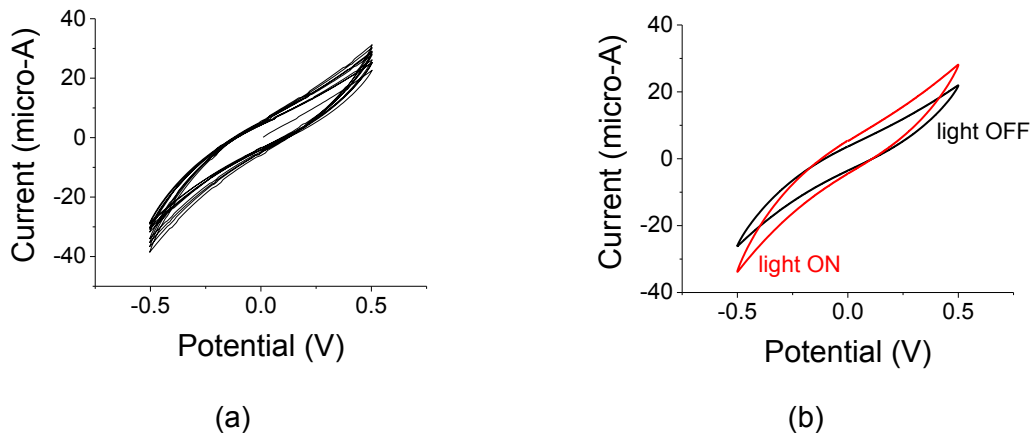

(a)  (b)

Fig. 7. CV curves for p-n like cell with IL electrolyte. Curves were obtained at a scan of 0.1 V/s.
(a) Calibration curve when the sample was heated when light was OFF from 23 to 35 °C.
(b) CV curves when the white light illumination was OFF and ON. Overall, the thermal effect from the IL dominated the capacitance increase.

### III.c4: p-n binders with aqueous solution

The thermal response of p-n like cells may be corroborated by using Al electrodes, 1 M $Na_2SO_4$ electrolyte and eliminating the n-Si QDs from the electrode composition. Fig. 8 shows CD curves for p(AC/PVP/n-Si)-n(AC/PEI) sample. The sample was heated by a 75 W lamp (although the Al current collector prevented any light penetration) and the temperature was recorded at the sample surface by using a thermocouple point probe. The AC was manufactured by GCC. The <2% difference in the discharge time may be more to do with the instability of the binder in the aqueous electrolyte than with the heat. Therefore, we reiterate that aqueous solutions at the small temperature range studied have little thermal effect on the cell capacitance.

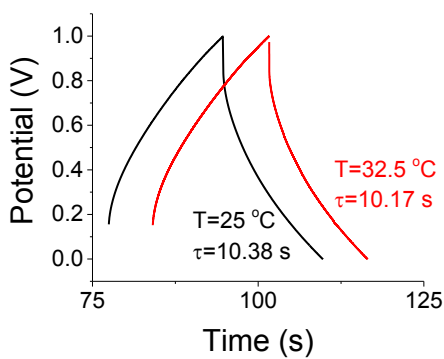

Fig. 8. CD curves at two temperatures for S-C, p(AC/PVP/n-Si)-n(AC/PEI) interfaced with Al electrodes and $Na_2SO_4$. The decay times for the discharge are shown in the figure. No major capacitance change is noted when using the aqueous solution.

### III.d. The effect of IR absorption band at 840 nm

As discussed earlier, incorporation of n-Si dots bleached the absorption band at 840 nm. Further corroboration may be obtained when measuring the capacitance while using a yellow optical filter. The filter transmits wavelengths larger than the cut-off wavelength of 550 nm, thus, eliminating the blue band from the white-light spectra. One has to factor the reduced overall light intensity when the light source is interfaced with such a filter. The intensity of the light source interfaced with the filter was 70% of the total white light output (both visible and IR) as measured by use of a bolometer. Thus, if the intensity related effects are linear, the capacitance increase due to the optical effect and/or thermal effect would be 70% of the effect without the filter.

Fig. 9a shows that a p(AC/PVP/n-Si)-n(AC/PEI) on FTO with IL electrolyte, illuminated with a filter-interfaced white-light source exhibited the same characteristics as non-illuminated S-C. On the other hand, a p(AC/PVP)-n(AC/PEI) on FTO (without n-Si) in Fig. 9b showed that the characteristics of light ON with, and without the filter are the same; both were different than the light OFF situation. When using the filter we eliminated the heating through blue light absorption but absorption was still present through the IR band.

All of the above may be summed up as follows: the IL was heated up through only the blue band absorption when the electrode contained n-Si; it was heated through both the blue band and the IR band in the absence of n-Si. In both cases, there is an increase in the relative cell capacitance.

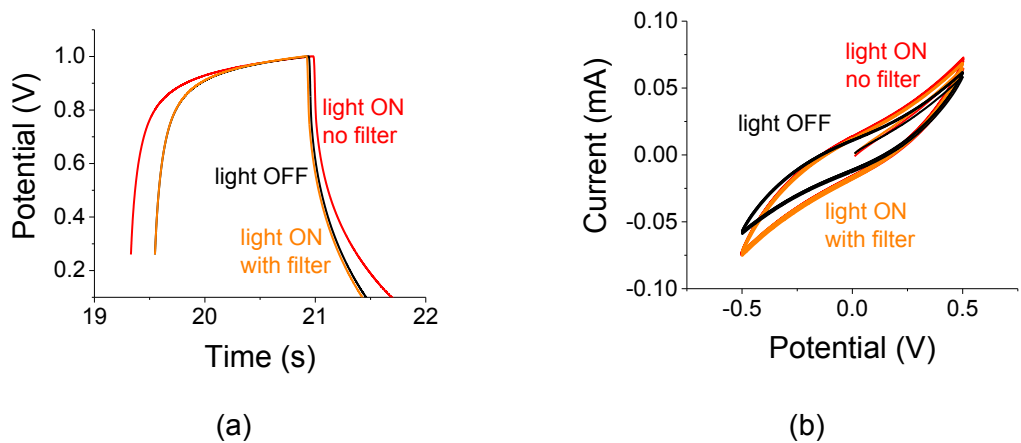

Fig. 9. Assessing the effect of the IR absorption band with IL electrolyte: (a) for p-n like S-C when n-Si are embedded in the 'p-like' electrode. Results for light OFF and light ON with the yellow filter are similar. (b) p-n only (no n-Si): results for light ON and light ON with the yellow filter are similar. The filter transmits wavelengths larger than the cut-off wavelength of 550 nm.

**Conclusions**

Asymmetric S-C, embedded with n-Si QD have shown a large capacitance increase. For aqueous cells it was due to local polarizations even in the absence of n-Si dots. For cells interfaced with IL electrolytes the capacitance increase was mostly due to optically-induced thermal effects in the electrolyte. Such elements exhibit promise for future novel, optically controlled supercapacitors.

1. Patrice Simon and Andrew Burke, "Nanostructured Carbons: Double-Layer Capacitance and More", The Electrochemical Society Interface, Spring 2008, 38-43.

2. Michio Inagaki, Hidetaka Konno, Osamu Tanaike, "Carbon materials for electrochemical capacitors", Journal of Power Sources 195 (2010) 7880–7903.

3. Zhang, S and Pan, N, "Supercapacitors Performance Evaluation", 2015. DOI:10.1002/aenm.201401401. https://escholarship.org/uc/item/26r5w8nc

4. Emre O. Polat and Coskun Kocabas, "Broadband Optical Modulators Based on Graphene Supercapacitors", Nano Lett. 2013, 13, 5851−5857. dx.doi.org/10.1021/nl402616t.

5. Xin Miao, Roberto Rojas-Cessa, Ahmed Mohamed and Haim Grebel, "The Digital Power Networks: Energy Dissemination Through a Micro-Grid", IEEE IoT, 2018

6. Roberto Rojas-Cessa, Haim Grebel, Zhengqi Jiang, Camila Fukuda, Henrique Pita, Tazima S. Chowdhury, Ziqian Dong and Yu Wan, "Integration of Alternative Energy Sources into Digital Micro-Grids", Environmental Progress & Sustainable Energy, 37, (2018) 155-164. DOI 10.1002/ep.

7. Jeffrey W. Long, Daniel Bélanger, Thierry Brousse, Wataru Sugimoto, Megan B. Sassin, and Olivier Crosnier, "Asymmetric electrochemical capacitors—Stretching the limits of aqueous electrolytes" MRS BULLETIN • VOLUME 36 • JULY 2011.

8. Yuanlong Shao, Maher F. El-Kady, Jingyu Sun, Yaogang Li, Qinghong Zhang, Meifang Zhu, Hongzhi Wang, Bruce Dunn, and Richard B. Kaner, "Design and Mechanisms of Asymmetric Supercapacitors", Chem. Rev. 2018, 118, 9233−9280. DOI: 10.1021/acs.chemrev.8b00252.

9. Mohammad S. Rahmanifar, Maryam Hemmati, Abolhassan Noori, Maher F. El-Kady, Mir F. Mousavi, Richard B. Kaner, "Asymmetric supercapacitors: An alternative to activated carbon negative electrodes based on earth abundant elements", Materials Today Energy 12 (2019) 26-36. https://doi.org/10.1016/j.mtener.2018.12.006

10. Anton V. Volkov, Hengda Sun, Renee Kroon, Tero-Petri Ruoko, Canyan Che, Jesper Edberg, Christian Müller, Simone Fabiano, and Xavier Crispin, "Asymmetric Aqueous Supercapacitor Based on p- and n‑Type Conducting Polymers", ACS Appl. Energy Mater. 2019, 2, 5350−5355. DOI: 10.1021/acsaem.9b00853.

11. Wen Shang, Yongtao Tan, Lingbin Kong, and Fen Ran, "Fundamental Triangular Interaction of Electron Trajectory Deviation and P−N Junction to Promote Redox Reactions for the High-EnergyDensity Electrode", ACS Appl. Mater. Interfaces 2020, 12, 29404−29413. https://dx.doi.org/10.1021/acsami.0c08299

12. Marilena Carbone, Mauro Missori, Laura Micheli, Pietro Tagliatesta and Elvira Maria Bauer, "NiO Pseudocapacitance and Optical Properties: Does The Shape Win?", Materials 2020, 13, 1417; doi:10.3390/ma13061417.

13. C. Wan, L. Yuan, and H. Shen, Int. J. Electrochem. Sci., Effects of Electrode Mass-loading on the Electrochemical Properties of Porous $MnO_2$ for Electrochemical Supercapacitor, 9 (2014) 1.

14. Yudong Li, Xianzhu Xu, Yanzhen He, Yanqiu Jiang and Kaifeng Lin, "Nitrogen Doped Macroporous Carbon as Electrode Materials for High Capacity of Supercapacitor", Polymers 2017, 9, 2; doi:10.3390/polym9010002.

15. Frederic Sauvage, Sarine Chhor, Arianna Marchioro, Jacques-E. Moser, and Michael Graetzel, "Butyronitrile-Based Electrolyte for Dye-Sensitized Solar Cells", J. Am. Chem. Soc. 2011, 133, 13103–13109. dx.doi.org/10.1021/ja203480w. DOI: 10.1021/cr400675m

16. Jihuai Wu, Zhang Lan, Jianming Lin, Miaoliang Huang, Yunfang Huang, Leqing Fan, and Genggeng Luo, "Electrolytes in Dye-Sensitized Solar Cells", Chem. Rev. 2015, 115, 2136−2173.

17. M. J. O'Connell, P. Boul, L. M. Ericson, C. Huffman, Y. Wang, E. Haroz, C. Kuper, J. Tour, K. D. A. and R. E. Smalley, Chem. Phys. Lett., Reversible water solubilization of single wall carbon-nanotubes by polymer wrapping, 342 (2001) 265.

18. M. Shim, A. Javey, N. W. S. Kam and H. Dai, J. Am. Chem. Soc., Polymer Functionalization for Air-Stable n-Type Carbon Nanotube Field-Effect Transistors, 123 (2001) 11512.

19. A. L. M. Reddy, M. M. Shaijumon, S. R. Gowda, and P. M. Ajayan, Multisegmented Au-MnO2/Carbon Nanotube Hybrid Coaxial Arrays for High-Power Supercapacitor Applications, J. Phys. Chem., 114 (2009) 658.

20. M. Kaempgen, C. K. Chan, J. Ma, Y. Cui, and G. Gruner, Nano Letts., Printable thin film supercapacitors using single-walled carbon nanotubes, 9 (2009) 1872.

21. Tazima S. Chowdhury and Haim Grebel, "Supercapacitors with electrical gates", Electrochimica Acta, 307, 459-464 (2019). https://doi.org/10.1016/j.electacta.2019.03.222

22. Tazima S. Chowdhury and Haim Grebel, "Ion-Liquid Based Supercapacitors with Inner Gate Diode-Like Separators", ChemEngineering 2019, 3(2), 39. https://doi.org/10.3390/chemengineering3020039

23. Balducci, Andrea and Dugas, R. and Taberna, Pierre-Louis and Simon, Patrice and Plee, Dominique and Mastragostino, Marina and Passerini, Stefano (2007) High temperature carbon–carbon supercapacitor using ionic liquid as electrolyte. Journal of Power Sources, 165, 922-927. ISSN 0378-7753

24. V. Ruiz, T. Huynh, S. R. Sivakkumar and A. G. Pandolfo, "Ionic liquid–solvent mixtures as supercapacitor electrolytes for extreme temperature operation", RSC Advances, 2012, 2, 5591–5598. DOI: 10.1039/c2ra20177a

25. Keh-Chyun Tsay, Lei Zhang, Jiujun Zhang, "Effects of electrode layer composition/thickness and electrolyte concentration on both specific capacitance and energy density of supercapacitor, "Electrochimica Acta 60 (2012) 428–436.